\documentclass[12pt]{article}
\usepackage{graphicx}
\usepackage[hypertex]{hyperref}

\begin{document}
\baselineskip 0.6cm

\def\simgt{\mathrel{\lower2.5pt\vbox{\lineskip=0pt\baselineskip=0pt
           \hbox{$>$}\hbox{$\sim$}}}}
\def\simlt{\mathrel{\lower2.5pt\vbox{\lineskip=0pt\baselineskip=0pt
           \hbox{$<$}\hbox{$\sim$}}}}

\begin{titlepage}

\begin{flushright}
UCB-PTH 07/16 \\
\end{flushright}

\vskip 2.2cm

\begin{center}

{\Large \bf
More Visible Effects of the Hidden Sector
}

\vskip 1.0cm

{\large Hitoshi Murayama, Yasunori Nomura, and David Poland}

\vskip 0.4cm

{\it Department of Physics, University of California,
     Berkeley, CA 94720, USA} \\
{\it Theoretical Physics Group, Lawrence Berkeley National Laboratory,
     Berkeley, CA 94720, USA} \\

\vskip 1.2cm

\abstract{There is a growing appreciation that hidden sector dynamics 
 may affect the supersymmetry breaking parameters in the visible sector 
 (supersymmetric standard model), especially when the dynamics is 
 strong and superconformal.  We point out that there are effects 
 that have not been previously discussed in the literature.  For 
 example, the gaugino masses are suppressed relative to the gravitino 
 mass.  We discuss their implications in the context of various 
 mediation mechanisms.  The issues discussed include anomaly mediation 
 with singlets, the $\mu$ ($B\mu$) problem in gauge and gaugino 
 mediation, and distinct mass spectra for the superparticles that 
 have not been previously considered.}

\end{center}
\end{titlepage}

\section{Introduction}
\label{sec:intro}

Supersymmetry has been widely recognized as an excellent solution to 
the hierarchy problem, as long as the superparticle masses are below 
the TeV scale.  However, such low scale supersymmetry is in conflict 
with the data from flavor physics, unless the spectrum is (close 
to) that of minimal flavor violation, i.e., violation of the 
$U(3)^5$ flavor symmetry comes only from the standard model 
Yukawa couplings.  There are several promising ways to mediate 
supersymmetry breaking effects preserving this property, for 
example, gauge mediation~\cite{Dine:1981gu,Dine:1994vc}, anomaly 
mediation~\cite{Randall:1998uk,Giudice:1998xp}, and gaugino 
mediation~\cite{Kaplan:1999ac}.

One interesting possibility for the origin of a hierarchically 
small scale for supersymmetry breaking is dynamical supersymmetry 
breaking~\cite{Witten:1981nf}.  Supersymmetry breaking is triggered 
at low energies by nontrivial infrared gauge dynamics of the hidden 
sector, which is then transmitted to the supersymmetric standard 
model (SSM) sector through a mediation mechanism preserving flavor. 
Traditionally, the spectrum of the superparticles has been calculated 
using the SSM renormalization group equations below the scale of 
mediation.  There is, however, a growing appreciation that the 
dynamics of the hidden sector may affect the supersymmetry breaking 
parameters in the SSM sector through renormalization group evolution 
between the mediation scale and the scale where the hidden sector 
fields decouple.

One of the most drastic examples of hidden sector dynamics is 
conformal sequestering~\cite{Luty:2001jh}, which occurs when the 
hidden sector exhibits strong superconformal dynamics.  (For a 
discussion on the effects of the hidden sector outside of the 
conformal regime, see~\cite{Cohen:2006qc}.)  This achieves the 
suppression of certain (dangerous) local operators connecting 
the hidden sector and SSM sector fields, and helps one to mediate 
supersymmetry breaking in a flavor universal manner.  The construction 
is motivated by the AdS/CFT correspondence~\cite{Maldacena:1997re}. 
If the SSM sector is located on the ``ultraviolet brane'' of a 
truncated AdS space~\cite{Randall:1999ee}, while the hidden sector 
is on the ``infrared brane,'' the physical separation between the 
two sectors due to the AdS bulk can be interpreted in terms of 
conformal dynamics in four dimensions.  This helps us see that 
purely four-dimensional theories can achieve apparent sequestering 
due to the strong conformal dynamics of the hidden sector. 
This class of dynamics has been further discussed 
in~\cite{Dine:2004dv,Sundrum:2004un,Schmaltz:2006qs}.

In this paper, we point out that hidden sector conformal dynamics 
has additional effects on the SSM sector parameters that have not been 
discussed in the literature.  Namely, operators that are linear in a 
singlet field in the hidden sector are sequestered by the wavefunction 
renormalization factor relative to the gravitino mass.  Note that the 
authors of Ref.~\cite{Dine:2004dv} stated that these operators are 
not sequestered, which we do not agree with.  There are at least three 
immediate consequences of this observation.  (1) Anomaly mediation 
does not require the absence of singlets in the hidden sector, as 
the gaugino masses are sequestered and the anomaly mediated piece 
can dominate.  (2) Conformal hidden sector dynamics can make 
gravity mediated contributions more harmful in gauge and gaugino 
mediated models, depending on the dynamics.  (3) The $\mu$ ($B\mu$) 
problem in gauge and gaugino mediation can in principle be solved by 
strong conformal dynamics, although it requires certain assumptions on 
the hidden sector dynamics.  In addition, if the sequestering effects 
are sufficiently strong, we find very specific mass spectra for the 
superparticles that have not been discussed in the literature and can 
be tested at future experiments.

The organization of the paper is as follows.  In section~\ref{sec:mech}, 
we provide a general discussion on the effects of a strong hidden 
sector on local operators connecting the hidden and SSM sector fields. 
Section~\ref{sec:conseq} summarizes the consequences of these effects 
on the SSM parameters.  In section~\ref{sec:gauge} we discuss possible 
scenarios in which the dominant mediation mechanism is gauge mediation. 
It is shown that strong conformal dynamics can provide a solution 
to the $\mu$ ($B\mu$) problem and/or lead to distinct spectra for 
the superparticles.  Gaugino and anomaly mediation are considered 
in sections~\ref{sec:gaugino} and \ref{sec:anomaly}.  Finally, 
discussion and conclusions are given in section~\ref{sec:concl}.

\section{General Discussions}
\label{sec:mech}

In this section, we present general discussions on the renormalization 
of operators that couple the hidden and SSM sector fields due to 
strong conformal dynamics in the hidden sector.

In many models of supersymmetry breaking, there are both gauge 
non-singlet and singlet fields in the hidden sector.  We generically 
call them $q$ and $S$, respectively, without referring to particular 
models.  They may be ``elementary'' or ``composite,'' but this 
distinction is not very clear in superconformal theories as they may 
allow for several inequivalent descriptions (duality).  To keep the 
discussion uniform, we always take the normalization such that these 
fields have mass dimension $+1$.%
\footnote{For example, the meson field $\bar{Q} Q$ in supersymmetric 
 QCD naturally has mass dimension $+2$, while we normalize it as 
 $S = \bar{Q} Q/\Lambda_*$, with $\Lambda_*$ being the strong scale.}
Note, however, that we are interested in models where the $q$ and 
$S$ fields participate in strong conformal dynamics, and hence their 
scaling properties are not dictated by their classical dimensions 
but rather their conformal dimensions.  We will generically refer 
to the chiral superfields of the SSM sector as $\phi$.

The direct couplings between the hidden and SSM sector fields can 
come in various local operators.  They are all higher dimension 
operators and suppressed by some energy scale $M$.  In gauge mediation 
models, it is related (but not necessarily equal to) the messenger 
scale.  In anomaly and gaugino mediated models, it is (generically) 
close to the Planck scale.

One class of direct interaction operators is quadratic in the hidden 
sector fields.  For example, operators that contribute to the scalar 
squared masses are
\begin{equation}
  {\cal O}_\phi : \qquad
  \int\!d^4\theta\, c^q_\phi \frac{q^\dagger q}{M^2} \phi^\dagger \phi,
\qquad
  \int\!d^4\theta\, c^S_\phi \frac{S^\dagger S}{M^2} \phi^\dagger \phi.
\label{eq:scalar-op}
\end{equation}
Other operators of interest are
\begin{equation}
  {\cal O}_{B\mu} : \qquad
  \int\!d^4\theta\, c^q_{B\mu} \frac{q^\dagger q}{M^2} H_u H_d 
  + {\rm h.c.},
\qquad
  \int\!d^4\theta\, c^S_{B\mu} \frac{S^\dagger S}{M^2} H_u H_d 
  + {\rm h.c.},
\label{eq:Bmu-op}
\end{equation}
that contribute to the $B\mu$ parameter (the holomorphic supersymmetry 
breaking mass squared) in the Higgs sector.  Here and below, the 
coefficients $c$'s are dimensionless.

Using the singlet fields, we can also consider operators linear in 
the hidden sector fields.  The gaugino mass operator is
\begin{equation}
  {\cal O}_\lambda: \qquad
  \int\!d^2\theta\, c^S_\lambda \frac{S}{M} 
    {\cal W}^{a\alpha} {\cal W}^a_\alpha + {\rm h.c.},
\label{eq:gaugino-op}
\end{equation}
where ${\cal W}^a_\alpha$ ($a=1,2,3$) are the field strength superfields 
for the standard model gauge group. The operators
\begin{equation}
  {\cal O}_A: \qquad
  \int\!d^4\theta\, c^S_A \frac{S}{M} \phi^\dagger \phi + {\rm h.c.}.
\label{eq:A-op}
\end{equation}
contribute to the $A$ and $B$ parameters (the parameters associated 
with holomorphic supersymmetry breaking scalar trilinear and bilinear 
interactions), as well as the scalar masses $|A|^2$.  Finally, the 
operator
\begin{equation}
  {\cal O}_\mu: \qquad
  \int\!d^4\theta\, c^S_\mu \frac{S^\dagger}{M} H_u H_d + {\rm h.c.},
\label{eq:mu-op}
\end{equation}
contributes to the $\mu$ parameter (the supersymmetric Higgs mass).%
\footnote{In principle, one may also consider direct superpotential 
 couplings between the hidden and SSM sector fields, such as 
 $\int\!d^2\theta\, S H_u H_d$ or $\int\!d^2\theta\, S Q_i U_j H_u/M_*$. 
 We assume their absence throughout the paper.}

Note that we have used the formalism of global supersymmetry in the 
above expressions.  This is sufficient for the purpose of discussing 
operators that arise from integrating out a set of messenger fields, 
e.g., gauge mediation.  Later, we will discuss gravity and anomaly 
mediated contributions, which require a formulation with local 
supersymmetry.  The terms integrated over a half of the superspace 
above will then include the conformal compensator field $\Phi$ as 
$\int\! d^2\theta\, \Phi^3$, while the terms over the full superspace 
as $\int\! d^4\theta\, \Phi^\dagger \Phi$~\cite{Cremmer:1978hn}. 
The latter should be regarded not as a part of the K\"ahler potential 
$K$, but rather the superspace density $f = -3 M_{\rm Pl}^2\, 
e^{-K/3M_{\rm Pl}^2}$ before the Weyl scaling that removes the field 
dependence in the Planck scale.  Here, $M_{\rm Pl}$ is the reduced 
Planck scale.  After the Weyl scaling, each chiral superfield needs 
to be further rescaled by $1/\Phi$ to obtain the usual kinetic terms, 
leaving a nontrivial $\Phi$ dependence in the various mass parameters. 
In vacua with supersymmetry breaking and no cosmological constant, 
$\Phi = 1 + \theta^2 m_{3/2}$, where $m_{3/2}$ is the gravitino mass. 
As we continue our discussion, it should be understood that there is 
an implicit compensator dependence in all of the mass parameters, and 
that any sequestering effects are occurring in $f$, and not in $K$.

In many cases, some of the operators Eqs.~(\ref{eq:scalar-op}~--%
~\ref{eq:mu-op}) are unwanted.  The operators ${\cal O}_\phi$ in 
Eq.~(\ref{eq:scalar-op}) and ${\cal O}_A$ in Eq.~(\ref{eq:A-op}) 
are potential sources of flavor changing neutral currents.  All 
of them are potential sources of $CP$ violation.  Both of these 
are constrained tightly by the data.  The purpose of conformal 
sequestering, then, is to help suppress any unwanted operators.

The main point is that, as long as the relevant fixed point is 
infrared attractive, conformal field theories can help achieve 
this suppression.  To see this, we can regard the SSM sector fields 
as background fields, and rescale the hidden sector fields to absorb 
the operators ${\cal O}_{\phi,B\mu}$ in Eqs.~(\ref{eq:scalar-op},%
~\ref{eq:Bmu-op}) into coupling constants of the theory.  As long as 
the fixed point is stable against deformations of the dimensionless 
coupling constants, the coupling constants flow to their infrared 
fixed point values by power laws, losing ``memory'' of the initial 
conditions.  Therefore, the unwanted operators can be suppressed by 
powers of energy scales.  If we can suppress all unwanted operators 
by power laws, while at the same time keeping those we need, the 
conformal sequestering is a success.

Most of the discussions on conformal sequestering so far have focused 
on the operators quadratic in the hidden sector fields.  However, 
it is important to consider operators linear in the hidden sector 
fields as well.  To the best of the current authors' knowledge, 
the only paper that has addressed this class of operators is 
Ref.~\cite{Dine:2004dv}.  They stated that this class of operators 
is not suppressed relative to the gravitino mass.  This observation 
would have allowed for an easy solution to the $B\mu$ problem in 
gauge mediation, since the unwanted operator ${\cal O}_{B\mu}$ of 
Eq.~(\ref{eq:Bmu-op}) would then be power suppressed at low energies 
while keeping the necessary operators ${\cal O}_{\lambda,\mu}$ of 
Eqs.~(\ref{eq:gaugino-op},~\ref{eq:mu-op}) (see section~\ref{sec:gauge} 
for more detail).  Unfortunately, we disagree with this statement. 
We instead find that the conformal sequestering is more complete 
than what they suggested; operators linear in the hidden sector 
fields are also suppressed relative to the gravitino mass.

To make the discussion more concrete, let us introduce a couple of 
energy scales.  We already defined $M$ as the scale appearing in the 
higher dimension operators that couple the hidden and SSM sector fields. 
This may be close to the Planck scale for anomaly or gaugino mediated 
supersymmetry breaking, or it may be a combination of energy scales 
in general, such as in gauge mediated supersymmetry breaking.  We 
also define the energy scale $\Lambda_*$ as the scale where the hidden 
sector enters into the conformal regime.

Since $S$ is singlet under the hidden sector gauge group, the 
superconformal algebra requires that it must have an $R$ charge 
greater than $2/3$ to preserve unitarity~\cite{Mack:1975je}. 
The anomalous dimension is given in terms of the $R$ charge by 
$3R/2-1$, and hence the wavefunction renormalization factor
\begin{equation}
  {\cal L} = \int\!d^4\theta\, Z_{S}(\mu_R)\, S^\dagger S,
\label{eq:kin-S}
\end{equation}
always satisfies
\begin{equation}
  Z_S(\mu_R) = \left(\frac{\Lambda_*}{\mu_R}\right)^{3R(S)-2} > 1,
\label{eq:Z-S}
\end{equation}
for $\mu_R < \Lambda_*$, where $\mu_R$ is the renormalization scale, 
$R(S)$ the $R$ charge of $S$, and we have taken $Z_{S}(\Lambda_*) = 1$. 
There are no 1PI diagrams that renormalize operators linear in $S$, 
and hence ${\cal O}_\lambda$ in Eq.~(\ref{eq:gaugino-op}), ${\cal O}_A$ 
in Eq.~(\ref{eq:A-op}), and ${\cal O}_\mu$ in Eq.~(\ref{eq:mu-op}) 
receive only the wavefunction renormalization $Z_S^{-1/2}(\mu_R)$. 
Note that this effect is always a {\it suppression} of the operators. 
Therefore, their respective forms at the energy scale $\mu_R \ll 
\Lambda_*$ are
\begin{equation}
  \int\!d^2\theta\, Z_S^{-1/2}(\mu_R)\, c^S_\lambda 
    \frac{S}{M} {\cal W}^{a\alpha} {\cal W}^a_\alpha + {\rm h.c.},
\label{eq:gaugino-op-Z}
\end{equation}
for the gaugino masses,
\begin{equation}
  \int\!d^4\theta\, Z_S^{-1/2}(\mu_R)\, c^S_A 
    \frac{S}{M} \phi^\dagger \phi + {\rm h.c.},
\label{eq:A-op-Z}
\end{equation}
for the $A$, $B$ parameters, and the $|A|^2$ part of the scalar 
squared masses, and
\begin{equation}
  \int\!d^4\theta\, Z_S^{-1/2}(\mu_R)\, c^S_\mu 
    \frac{S^\dagger}{M} H_u H_d + {\rm h.c.},
\label{eq:mu-op-Z}
\end{equation}
for the $\mu$ parameter.

The $S$ field acquires an $F$-component vacuum expectation value (VEV) 
if there is a linear term in the superpotential, i.e., if there is an 
operator
\begin{equation}
  \int\!d^2\theta\, f^2 S + {\rm h.c.},
\label{eq:S-linear}
\end{equation}
where $f$ has mass dimension one.  In the basis where the $S$ field 
is canonically normalized, this linear term is also suppressed in 
the infrared as
\begin{equation}
  \int\!d^2\theta\, Z_S^{-1/2}(\mu_R)\, f^2 S + {\rm h.c.}.
\label{eq:S-linear-can}
\end{equation}
The $F$-component VEV for the canonically normalized $S$ is
\begin{equation}
  F_S = -Z_S^{-1/2}(\mu_R)\, f^{* 2},
\label{eq:F_S-can}
\end{equation}
and the vacuum energy $V_0 = |Z_S^{-1/2}(\mu_R)\, f^2|^2$, and hence 
the gravitino mass is
\begin{equation}
  m_{3/2} \approx Z_S^{-1/2}(\mu_R) 
    \frac{|f|^2}{M_{\rm Pl}}.
\label{eq:m32}
\end{equation}
The apparent suppression $Z_S^{-1/2}(\mu_R)$ of Eq.~(\ref{eq:m32}), 
however, does not have much physical meaning, since it suppresses 
all the $\mu$ and supersymmetry breaking parameters equally.  For 
example, the gaugino masses are given by
\begin{equation}
  |M_a| \approx Z_S^{-1/2}(\mu_R)\, \frac{|c^S_\lambda F_S|}{M} 
  = Z_S^{-1}(\mu_R)\, 
    \frac{|c^S_\lambda f^2|}{M}.
\label{eq:Ma-Z}
\end{equation}
In this last expression, one factor of $Z_S^{-1/2}(\mu_R)$ comes 
from that of Eq.~(\ref{eq:m32}), but the other $Z_S^{-1/2}(\mu_R)$ 
from the suppression of the coefficient of Eq.~(\ref{eq:gaugino-op-Z}). 
It is this latter $Z_S^{-1/2}(\mu_R)$ that provides the suppression 
of the gaugino masses relative to the gravitino mass: $M_a/m_{3/2} 
\sim Z_S^{-1/2}(\mu_R)$.  Similar analyses also apply to the $\mu$ 
and $A$ parameters.  Therefore, the gaugino masses, $\mu$, and $A$ 
parameters receive a stronger suppression than the gravitino mass, 
affecting phenomenology and model building as we will discuss later.

In contrast to the operators linear in $S$, the operators 
${\cal O}_{\phi,B\mu}$ in Eqs.~(\ref{eq:scalar-op},~\ref{eq:Bmu-op}) 
receive corrections from 1PI diagrams in addition to the wavefunction 
renormalization factors.  Note that the $R$ charges of gauge non-singlet 
fields can be less than $2/3$ because they do not appear as asymptotic 
states, and hence the standard representation theory does not apply. 
To simplify the discussion, let us ignore operator mixing at this 
moment, and pretend that these operators renormalize by themselves. 
We then find
\begin{equation}
  \int\!d^4\theta\, \left( \frac{\mu_R}{\Lambda_*} \right)^{\alpha_q} 
    Z_q^{-1}(\mu_R)\, c^q_\phi \frac{q^\dagger q}{M^2} \phi^\dagger \phi,
\qquad
  \int\!d^4\theta\, \left( \frac{\mu_R}{\Lambda_*} \right)^{\alpha_S} 
    Z_S^{-1}(\mu_R)\, c^S_\phi \frac{S^\dagger S}{M^2} \phi^\dagger \phi,
\label{eq:scalar-op-Z}
\end{equation}
for the scalar squared masses, and
\begin{equation}
  \int\!d^4\theta\, \left( \frac{\mu_R}{\Lambda_*} \right)^{\alpha_q} 
    Z_q^{-1}(\mu_R)\, c^q_{B\mu} \frac{q^\dagger q}{M^2} H_u H_d 
    + {\rm h.c.},
\qquad
  \int\!d^4\theta\, \left( \frac{\mu_R}{\Lambda_*} \right)^{\alpha_S} 
    Z_S^{-1}(\mu_R)\, c^S_{B\mu} \frac{S^\dagger S}{M^2} H_u H_d 
    + {\rm h.c.},
\label{eq:Bmu-op-Z}
\end{equation}
for the $B\mu$ term, where $Z_q(\mu_R)$ is defined analogously to 
$Z_S(\mu_R)$; see Eq.~(\ref{eq:Z-S}).%
\footnote{If the operators ${\cal O}_{\phi,B\mu}$ in 
 Eqs.~(\ref{eq:scalar-op},~\ref{eq:Bmu-op}) are generated at a scale 
 $m_f < \Lambda_*$, the factors $(\mu_R/\Lambda_*)^{\alpha_{q,S}}$ 
 in Eqs.~(\ref{eq:scalar-op-Z},~\ref{eq:Bmu-op-Z}) should be replaced 
 by $(\mu_R/m_f)^{\alpha_{q,S}}$.}
The exponents $\alpha_q$ and $\alpha_S$ are common to the operators 
in Eqs.~(\ref{eq:scalar-op-Z}) and (\ref{eq:Bmu-op-Z}), since 
the dependence of these operators on the hidden sector fields 
is the same.  Note that here we defined $\alpha_{q,S}$ to 
parameterize the 1PI corrections; for example, if one of the 
operators in Eqs.~(\ref{eq:scalar-op-Z},~\ref{eq:Bmu-op-Z}) 
corresponds to a conserved current in the hidden sector, the 
$(\mu_R/\Lambda_*)^{\alpha_{q,S}}$ factor exactly cancels the 
wavefunction renormalization factor $Z_{q,S}^{-1}(\mu_R)$.

An interesting and often crucial question is the relative speed 
of suppression (sequestering) between the operators quadratic and 
linear in $S$.  Suppose that there is no mixing between operators 
quadratic in $S$ and those quadratic in $q$, and that only $S$ has 
a supersymmetry breaking VEV.  Then, if there were no extra exponent 
$\alpha_S$, all the $\mu$ and soft parameters would receive similar 
suppressions as $M_a \sim \mu \sim A \propto Z_S^{-1/2} F_S$ and 
$m_I^2 \sim B\mu \propto Z_S^{-1} F_S^2$, while $m_{3/2} \propto F_S$. 
Here, $m_I^2$ represent the supersymmetry breaking scalar squared 
masses.  Realistically, however, the situation is not that simple. 
The operators of the form ${\cal O}_\phi$ in Eq.~(\ref{eq:scalar-op}) 
(and ${\cal O}_{B\mu}$ in Eq.~(\ref{eq:Bmu-op})) in general mix 
with each other, and $\alpha_{q,S}$ are nonzero.  In this case, the 
suppression of the operators quadratic in $S$ is controlled by the 
smallest eigenvalue of the $2\gamma_i \delta_{ij} + \alpha_{ij}$ 
matrix, which we define as $2\gamma_S+\hat{\alpha}_S$.  Here, $i,j$ 
runs over $q$ and $S$, and $\gamma_q \equiv 3R(q)/2-1$ and $\gamma_S 
\equiv 3R(S)/2-1$ are the anomalous dimensions of the $q$ and 
$S$ fields.  (For a detailed discussion on operator mixing, see 
Appendix~\ref{app:op-mix}.)

One additional subtlety is that the operators quadratic in $S$ also 
mix in a calculable way with the operators linear in $S$. In particular, 
for the non-Higgs fields it is really the combination $c^S_\phi 
- |c^S_A|^2$ that ends up being suppressed by the exponent $2\gamma_S 
+ \hat{\alpha}_S$ (after potentially mixing with other quadratic 
operators), and this is the same combination of operators that 
contributes to the scalar squared masses.  Similarly, for the Higgs 
fields it is the combination $c^S_\phi - |c^S_A|^2 - |c^S_\mu|^2$ 
that ends up being suppressed by the same exponent, and this is the 
operator that contributes to $m_{H_{u,d}}^2 + \mu^2$.  Finally, the 
combination of operators that contributes to the $B\mu$ parameter, 
$c^S_{B\mu} - c^S_\mu (c^S_{A,H_u} + c^S_{A,H_d})$, is renormalized 
in the same way.

One can then obtain the qualitatively different outcomes:
\begin{equation}
\begin{array}{lll}
  \mbox{Case 1:} & M_a^2 \sim \mu^2 \sim A^2 
    \gg m_{Q_i,U_i,D_i,L_i,E_i}^2 \sim B\mu \sim m_{H_{u,d}}^2 + \mu^2 
    & (\hat{\alpha}_S > 0),
\\
  \mbox{Case 2:} & M_a^2 \sim \mu^2 \sim A^2 
    \ll m_{Q_i,U_i,D_i,L_i,E_i}^2 \sim B\mu \sim m_{H_{u,d}}^2 
    & (\hat{\alpha}_S < 0),
\end{array}
\end{equation}
depending on the sign of the exponent $\hat{\alpha}_S$.  (In the 
absence of the operator mixing, $\hat{\alpha}_S = \alpha_S$.)  In 
addition, since all the soft parameters are suppressed {\it relative 
to} the gravitino mass (except for those that correspond to conserved 
currents), it is also possible that they are all subdominant relative 
to the gravitino mass, in which case anomaly mediation may be 
dominant (Case~3).  Unfortunately, for a given strongly coupled 
conformal theory, it is not possible to work out the signs or 
magnitudes of the exponents $\alpha_{q,S}$ with the currently 
available technology.  We will therefore discuss all three cases 
on equal footing in the rest of the paper.

Note that we are only considering operators linear or quadratic in 
the hidden sector fields because they are the lowest dimension operators 
that contribute to the soft supersymmetry breaking parameters in the 
SSM sector.  However, due to the incalculable strong dynamics, we 
cannot exclude the possibility that even higher dimension operators, 
i.e., cubic, quartic, or beyond in the hidden sector fields, receive 
anomalously large enhancements relative to the lower dimension operators 
and become as important.  See section~\ref{sec:concl} for more on 
this point.

\section{Consequences on the SSM Parameters}
\label{sec:conseq}

What effect does the strong hidden sector renormalization, discussed 
in the previous section, have on the $\mu$ and supersymmetry breaking 
parameters in the SSM sector?  As we have seen, terms linear in the 
$S$ field are power suppressed in a way that is controlled exactly 
by the $R$ charge of $S$.  On the other hand, particular combinations 
of terms quadratic in the $S$ field and terms linear in the $S$ field 
are suppressed (assuming the fixed point is infrared attractive) 
by an incalculable amount, determined by the rate at which the theory 
flows back towards the conformal fixed point.  Rather generically, 
if the effects are strong, we expect that one of these classes of 
operators will completely dominate over the other.  An important point 
here is that the {\it relative} strengths of the operators linear in 
$S$ remain fixed, since they are all suppressed by the same amount. 
Similarly, the relative strengths of the operators quadratic in $S$ 
(in combination with linear operators) also do not change.%
\footnote{In the case that there is operator mixing and/or multiple 
 singlets with $F$-component VEVs, the relative strengths among 
 linear operators and/or quadratic operators can in principle change 
 depending on how they project onto the ``eigenvectors'' of the 
 renormalization group evolution.  These effects, however, are 
 typically of $O(1)$ if present, and do not affect the arguments 
 below.}

The operators linear in $S$, ${\cal O}_{\lambda,A,\mu}$ in 
Eqs.~(\ref{eq:gaugino-op},~\ref{eq:A-op},~\ref{eq:mu-op}), contribute 
to the gaugino masses $M_a$, $\mu$ parameter, scalar squared masses 
$m_I^2$, and $A$ and $B$ parameters.  Note, however, that because the 
scalar masses, $A$ parameters, and $B$ parameter are all generated 
by the single operator ${\cal O}_A$ in Eq.~(\ref{eq:A-op}), there 
are simple relations among them.  On the other hand, the operators 
quadratic in $S$, ${\cal O}_{\phi,B\mu}$ in Eqs.~(\ref{eq:scalar-op},%
~\ref{eq:Bmu-op}), also independently contribute to the scalar masses 
and $B\mu$ term, and the dynamics may actually drive these to cancel 
the contributions from the linear operators.  Thus, if the hidden 
sector effects are strong, we are generically led to one of the 
the following situations:

\subsection*{Case 1: Linear operator dominance}
\label{subsec:case-1}

In this case, any initial conditions in the quadratic operators 
are suppressed, and they are dynamically driven to cancel out the 
contributions to the soft parameters from the linear operators. 
As long as these linear operators are all generated at approximately 
the same size, we obtain the following spectrum at the scale where 
the hidden sector exits from the conformal fixed point:
\begin{eqnarray}
   && m_{Q_i,U_i,D_i,L_i,E_i}^2 = 0,
\qquad
   m_{H_{u,d}}^2 = -\mu^2,
\nonumber\\
   && a_{IJK} = y_{IJK} (A_I + A_J + A_K),
\\
   && B = 0,
\nonumber
\end{eqnarray}
where $I,J,K$ runs over the SSM matter and Higgs fields, $Q_i, U_i, D_i, 
L_i, E_i, H_u, H_d$ ($i=1,2,3$), and $A_I$ represent the coefficients 
of the operators $\int\!d^4\theta\, S \phi_I^\dagger \phi_I$ times 
$F_S$, which are of the same order as the gaugino masses and the $\mu$ 
parameter
\begin{equation}
  M_a \,\approx\, \mu \,\approx\, A_I.
\end{equation}
The soft parameters $m_I^2$ and $a_{IJK}$ are defined by 
${\cal L}_{\rm soft} = - m_I^2 \phi_I^\dagger \phi_I - (a_{IJK} 
\phi_I \phi_J \phi_K + {\rm h.c.})$, and $y_{IJK}$ are the Yukawa 
couplings: $W = y_{IJK} \phi_I \phi_J \phi_K$.  Here, we have 
neglected, for simplicity, possible mixings between different 
generations in $A_I$, which may be present in general.

In order to avoid excessive flavor changing processes, the parameters 
$A_I$ must take a special form in flavor space.  One simple possibility 
is that the $A_I$ operators are generated only for the Higgs and third 
generation matter fields.  In this case, we obtain
\begin{equation}
  m_I^2 = 0,
\label{eq:12-mI2}
\end{equation}
\begin{equation}
  a_{IJH_u} = y_{IJH_u} A_{H_u},
\qquad
  a_{IJH_d} = y_{IJH_d} A_{H_d},
\label{eq:12-a}
\end{equation}
for the first two generation matter fields, and
\begin{equation}
  m_{Q_3,U_3,D_3,L_3,E_3}^2 = 0,
\qquad
  m_{H_{u,d}}^2 = -\mu^2,
\label{eq:3-mI2}
\end{equation}
\begin{equation}
  a_t = y_t (A_{Q_3}+A_{U_3}+A_{H_u}),
\qquad
  a_b = y_b (A_{Q_3}+A_{D_3}+A_{H_d}),
\qquad
  a_\tau = y_\tau (A_{L_3}+A_{E_3}+A_{H_d}),
\label{eq:3-a}
\end{equation}
\begin{equation}
  B = 0,
\label{eq:12-3-B}
\end{equation}
for the third generation matter and Higgs fields.  Here, $y_t, y_b, 
y_\tau$ are the top, bottom, and tau Yukawa couplings, $a_t, a_b, 
a_\tau$ the corresponding scalar trilinear interactions, and
\begin{equation}
  M_a \,\approx\, \mu \,\approx\, A_I,
\label{eq:12-3-comp}
\end{equation}
where $I=Q_3,U_3,D_3,L_3,E_3,H_u,H_d$.  A special case of this 
spectrum is obtained if only the Higgs fields have the $A_I$ 
operators: $A_{Q_3} = A_{U_3} = A_{D_3} = A_{L_3} = A_{E_3} = 0$.

The spectra given above represent the running parameters evaluated 
at the scale where the hidden sector exits from the conformal regime, 
which is generically much larger than the weak scale.  The low-energy 
superparticle masses are then obtained by evolving these parameters 
down to the weak scale using the renormalization group equations. 
Since the hidden sector already leaves the strong conformal regime, 
these evolutions are dominated by loops of the SSM states, i.e., the 
running is well approximated by the standard SSM renormalization 
group equations.

\subsection*{Case 2: Quadratic operator dominance}
\label{subsec:case-2}

In this case, the quadratic operators (or at least one of them in the 
case that there are operator mixings) are suppressed more slowly than 
the linear operators.  This can easily be the case, for example, if the 
quadratic operators contain a global symmetry current(s) of the hidden 
sector, which does not receive any suppression factor.  This leads to 
the split spectrum
\begin{eqnarray}
  m_I^2,\, B\mu \,\gg\, M_a^2,\, \mu^2,\, a_{IJK}^2,
\label{eq:case-2}
\end{eqnarray}
at the scale where the hidden sector exits the conformal regime. 
This splitting is preserved by renormalization group evolution at 
lower energies, so if the splitting is very large, the spectrum 
requires a severe fine-tuning in electroweak symmetry breaking.

The spectrum, however, does not require fine-tuning if the splitting 
is not very large.  The exact spectrum is determined by the mediation 
mechanism, and may show a distinct pattern which is not common to 
the scenarios in which the hidden sector dynamics are not taken into 
account.  If we take gauge mediation, for example, we obtain a somewhat 
interesting spectrum in which the number of messenger fields appears 
to be {\it fractional}, as we will see in section~\ref{sec:gauge}.

\subsection*{Case 3: Anomaly mediation dominance}
\label{subsec:case-3}

In both of the previous situations, it is worth emphasizing that 
{\it all} the parameters are being suppressed relative to the 
gravitino mass $m_{3/2} \approx F_S/M_{\rm Pl}$, except for operators 
corresponding to conserved currents of the hidden sector, which 
we assume to be absent here.  If the suppressions of both types of 
operators are strong enough, then, we will be led to the situation 
where the anomaly mediated contribution dominates.  It is also possible, 
depending on the amount of suppressions, that the dominant contributions 
to the SSM sector parameters come both from anomaly mediation and 
some of the local operators involving the $S$ field.  These points 
will be discussed further in section~\ref{sec:anomaly}.

\section{Gauge Mediation}
\label{sec:gauge}

In this section, we take gauge mediation as the dominant mediation 
mechanism generating the local operators in Eqs.~(\ref{eq:scalar-op}~%
--~\ref{eq:mu-op}), and consider the possible implications of the 
hidden sector dynamics discussed in the previous sections.  The 
situation is different depending on which of Case~1 or Case~2 is 
realized as a result of the hidden sector dynamics.  We first discuss 
the implications of Case~1 in subsection~\ref{subsec:mu-sol}, and 
then discuss those of Case~2 in subsection~\ref{subsec:fractional}. 
Finally, we discuss the competition with gravity and anomaly 
mediation in subsection~\ref{subsec:compet}.

\subsection{Solution to the {\boldmath $\mu$} ({\boldmath $B\mu$}) 
problem with conformal dynamics}
\label{subsec:mu-sol}

A major difficulty of the gauge mediation scenario is the so-called 
$\mu$ problem --- it is difficult to obtain phenomenologically acceptable 
values for the $\mu$ and $B\mu$ parameters.  In fact, a careful look 
at the problem shows that it is really a $B\mu$ problem, rather than 
a $\mu$ problem (see, e.g., \cite{Dvali:1996cu}).  In gauge mediation, 
the gaugino masses, $M_a$, and the scalar squared masses, $m_I^2$, 
arise at one and two loops, respectively, so that these masses 
have the comparable size $M_a \approx m_I \approx (g^2/16\pi^2) 
(F_{\rm mess}/M_{\rm mess})$, where $F_{\rm mess}/M_{\rm mess} 
\approx (10$~--~$100)~{\rm TeV}$ is the scale characterizing the 
strength of the mediation.  Now, it is not so difficult to come 
up with a model in which the $\mu$ term is generated at one loop, 
$\mu \approx (1/16\pi^2)(F_{\rm mess}/M_{\rm mess})$, so that it 
is comparable to the gaugino and scalar masses.  However, such 
a model also tends to generate the $B\mu$ term at one loop, 
$B\mu \approx (1/16\pi^2)(F_{\rm mess}/M_{\rm mess})^2$, leading 
to the parameter $B$ being one-loop {\it enhanced} relative to the 
other supersymmetry breaking masses, $B \equiv B\mu/\mu \approx 
F_{\rm mess}/M_{\rm mess}$.  Since the size of $B$ should be 
smaller than or of the order of the weak scale to obtain successful 
phenomenology, this is not acceptable.

We point out here that this problem can be solved if the hidden sector 
has strong conformal dynamics exhibiting the property described as 
Case~1 in section~\ref{sec:conseq}.  Suppose that gauge mediation 
arises due to vector-like messenger superfields $f, \bar{f}$ having 
a mass $m_f$ and a coupling to the hidden sector superfield $S$ in 
the superpotential~\cite{Murayama:2007fe}:
\begin{equation}
  W = - m_f \bar{f} f + \lambda S \bar{f} f.
\label{eq:W_mess}
\end{equation}
Here, $S$ is a superfield responsible for supersymmetry breaking, 
$\langle S \rangle = \theta^2 F_S$, and the coupling $\lambda$ 
encodes the information on the classical dimension of the (composite) 
operator $S$:
\begin{equation}
  \lambda = O\Biggl( \biggl(\frac{\Lambda_*}{M_*}\biggr)^{d_S-1} \Biggr).
\label{eq:lambda}
\end{equation}
Here, $M_*$ is the cutoff scale of the theory, which can be taken to 
be around the Planck scale $M_* \approx M_{\rm Pl}$, and $d_S$ the 
classical mass dimension of $S$.  The parameter $m_f$ can be taken 
real and positive without loss of generality.

At the scale $m_f$, the messenger fields are integrated out.  This 
generates the operators
\begin{equation}
  {\cal L} = \frac{1}{2} D_f \int\!d^2\theta\, 
    \sum_a \frac{\lambda(m_f)}{16\pi^2 m_f} S\, 
    {\cal W}^{a\alpha} {\cal W}^a_\alpha + {\rm h.c.},
\label{eq:gmsb-gaugino-op}
\end{equation}
where $a = 1,2,3$ represents the standard model gauge groups, 
${\cal W}^a_\alpha$ the corresponding field-strength superfields, and 
$D_f$ the Dynkin index of the messengers ($1$ for ${\bf 5}+{\bf 5}^*$, 
$3$ for ${\bf 10}+{\bf 10}^*$ etc), and
\begin{equation}
  {\cal L} = -D_f \int\!d^4\theta\, \sum_I \sum_a 
    \frac{2 g_a^4 C_I^a |\lambda(m_f)|^2}{(16\pi^2)^2 m_f^2} 
    S^\dagger S\, \phi_I^\dagger \phi_I,
\label{eq:gmsb-scalar-op}
\end{equation}
where $g_a$ are the standard model gauge couplings evaluated at 
$m_f$, and $C_I^a$ the quadratic Casimir coefficients.  Here, 
$\lambda(m_f)$ is the physical coupling $\lambda$ evaluated at 
$\mu_R \approx m_f$:
\begin{equation}
  \lambda(m_f) 
  = \left( \frac{m_f}{\Lambda_*} \right)^{\gamma_S} \lambda.
\label{eq:lambda-mf}
\end{equation}
After $S$ acquires a supersymmetry breaking VEV, the operators 
of Eqs.~(\ref{eq:gmsb-gaugino-op}) and (\ref{eq:gmsb-scalar-op}) 
become the gaugino masses and scalar squared masses, respectively.

In order to solve the $\mu$ problem, the $\mu$ parameter must be 
generated with a size comparable to the gaugino masses.  It is, in fact, 
not very difficult to generate the $\mu$ term also at one loop as
\begin{equation}
  {\cal L} \approx \int\!d^4\theta\, \frac{\lambda(m_f)^*}{16\pi^2 m_f} 
    S^\dagger\, H_u H_d + {\rm h.c.},
\label{eq:gmsb-mu-op}
\end{equation}
(for an example of such models, see Appendix~\ref{app:mu}).  This leads 
to a $\mu$ parameter of the same order as the gaugino and scalar masses 
generated by Eqs.~(\ref{eq:gmsb-gaugino-op},~\ref{eq:gmsb-scalar-op}):
\begin{equation}
  M_a \approx m_I \approx \mu 
    \approx \left| \frac{\lambda(m_f) F_S}{16\pi^2 m_f} \right|.
\label{eq:mf-soft}
\end{equation}
The problem is that in any simple models producing the operator 
Eq.~(\ref{eq:gmsb-mu-op}) at one loop, the same one-loop diagram 
also generates another operator
\begin{equation}
  {\cal L} \approx \int d^4 \theta \frac{|\lambda(m_f)|^2}{16\pi^2 m_f^2} 
    S^\dagger S\, H_u H_d + {\rm h.c.},
\label{eq:gmsb-Bmu-op}
\end{equation}
which leads to a large $B$ parameter
\begin{equation}
  B \equiv \frac{B\mu}{\mu} 
    \approx \left| \frac{\lambda(m_f) F_S}{m_f} \right| 
    \gg M_a, m_I, \mu.
\label{eq:mf-B}
\end{equation}
This is nothing but the $B\mu$ problem in gauge mediation discussed 
earlier.  Models that generate the operator Eq.~(\ref{eq:gmsb-mu-op}) 
at one loop also typically generate the operators
\begin{equation}
  {\cal L} \approx \int\!d^4\theta\, 
    \left( \frac{\lambda(m_f)}{16\pi^2 m_f} S\, H_u^\dagger H_u 
         + \frac{\lambda(m_f)}{16\pi^2 m_f} S\, H_d^\dagger H_d 
      + {\rm h.c.} \right),
\label{eq:gmsb-AH-op}
\end{equation}
which contribute to $A_{H_u}$, $A_{H_d}$, $B$, $m_{H_u}^2$ and 
$m_{H_d}^2$.  These operators, however, are harmless, since the 
generated soft masses are of the same order as the gaugino masses, 
and the $A$ terms induced preserve flavor, i.e., $a_{IJK}$ are 
proportional to the Yukawa matrices, $y_{IJK}$, in flavor space.

The operators of Eqs.~(\ref{eq:gmsb-gaugino-op},~\ref{eq:gmsb-scalar-op},%
~\ref{eq:gmsb-mu-op},~\ref{eq:gmsb-Bmu-op},~\ref{eq:gmsb-AH-op}) are 
the ones generated at $m_f$ and relevant for the $\mu$ and supersymmetry 
breaking parameters in the SSM sector.  They lead to an unacceptably 
large $B$ parameter.  Note, however, that these correspond to the 
$\mu$ and supersymmetry breaking masses evaluated at the scale $m_f$. 
If the hidden sector interactions are strong below the scale $m_f$ 
down to some scale $m_X$ where conformality is broken, as we are 
assuming here, then the operators of Eqs.~(\ref{eq:gmsb-gaugino-op},%
~\ref{eq:gmsb-scalar-op},~\ref{eq:gmsb-mu-op},~\ref{eq:gmsb-Bmu-op},%
~\ref{eq:gmsb-AH-op}) receive strong renormalization effects in the 
energy interval between $m_f$ and $m_X$.  The pattern of the soft masses 
at $m_X$ ($\ll m_f$) depends on which of Case~1 and Case~2 is realized, 
and here we assume that Case~1 is realized.  In this case, the operators 
Eqs.~(\ref{eq:gmsb-scalar-op},~\ref{eq:gmsb-Bmu-op}) (in particular 
combinations with the linear operators) are damped compared with 
the operators Eqs.~(\ref{eq:gmsb-gaugino-op},~\ref{eq:gmsb-mu-op},%
~\ref{eq:gmsb-AH-op}), with the relative strengths of the operators 
Eqs.~(\ref{eq:gmsb-gaugino-op}), (\ref{eq:gmsb-mu-op}) and 
(\ref{eq:gmsb-AH-op}) preserved.

The $\mu$ and supersymmetry breaking parameters at $m_X$ then satisfy 
the pattern of Eqs.~(\ref{eq:12-mI2}~--~\ref{eq:12-3-comp}) with 
$A_{Q_3,U_3,D_3,L_3,E_3} = 0$:
\begin{equation}
  m_{Q_i,U_i,D_i,L_i,E_i}^2 = 0,
\label{eq:model-mI2}
\end{equation}
\begin{equation}
  (a_u)_{ij} = (y_u)_{ij} A_{H_u},
\qquad
  (a_d)_{ij} = (y_d)_{ij} A_{H_d},
\qquad
  (a_e)_{ij} = (y_e)_{ij} A_{H_d},
\label{eq:model-a}
\end{equation}
\begin{equation}
  m_{H_u}^2 = -\mu^2,
\qquad
  m_{H_d}^2 = -\mu^2,
\qquad
  B = 0,
\label{eq:model-H}
\end{equation}
\begin{equation}
  M_a \,\approx\, \mu \,\approx\, A_{H_u} \,\approx\, A_{H_d},
\label{eq:model-comp}
\end{equation}
where $(y_u)_{ij}$, $(y_d)_{ij}$ and $(y_e)_{ij}$ are the up-type quark, 
down-type quark and charged lepton Yukawa matrices, and $(a_u)_{ij}$, 
$(a_d)_{ij}$ and $(a_e)_{ij}$ the corresponding scalar trilinear 
interactions.  The low-energy superparticle masses are obtained by 
evolving these parameters from $m_X$ down to the weak scale.  This 
evolution is well approximated by the standard SSM renormalization 
group equations.  Since both $\mu$ and $B$ at the weak scale are 
the same order as the other soft masses, the $\mu$ ($B\mu$) problem 
is solved.

The pattern in Eqs.~(\ref{eq:model-mI2}~--~\ref{eq:model-comp}) resembles 
that of gaugino mediation with a low compactification scale, or standard 
gauge mediation with a very large number of messenger fields.  These 
theories, however, lead to a Landau pole for the standard model gauge 
couplings below the unification scale, and thus are not compatible 
with perturbative gauge coupling unification.  Our theory is fully 
compatible with perturbative gauge coupling unification.  Moreover, 
the present scenario leads to particular relations for $m_{H_u}^2$, 
$m_{H_d}^2$, $\mu$ and $B$, which can be tested at future collider 
experiments.

\subsection{Spectrum with a fractional number of messenger fields}
\label{subsec:fractional}

We now consider the case that the hidden sector exhibits the dynamics 
of Case~2, rather than Case~1.  This happens, for example, if one or 
more of the $S^\dagger S$ operators corresponds to a conserved global 
current(s) of the hidden sector dynamics.  In this case, renormalization 
of the operators Eqs.~(\ref{eq:gmsb-gaugino-op},~\ref{eq:gmsb-scalar-op},%
~\ref{eq:gmsb-mu-op},~\ref{eq:gmsb-Bmu-op},~\ref{eq:gmsb-AH-op}) below 
$m_f$ is different from that discussed in the previous subsection. 
Specifically, at the scale $m_X$ where the hidden sector leaves the 
conformal fixed point, the operators Eqs.~(\ref{eq:gmsb-gaugino-op},%
~\ref{eq:gmsb-mu-op},~\ref{eq:gmsb-AH-op}) are suppressed compared 
with Eqs.~(\ref{eq:gmsb-scalar-op},~\ref{eq:gmsb-Bmu-op}).  This 
leads to a split spectrum
\begin{equation}
  m_I^2 \,\approx\, B\mu \,\gg\, 
    M_a^2 \,\approx\, \mu^2 \,\approx\, A_{H_u}^2 \,\approx\, A_{H_d}^2,
\label{eq:split}
\end{equation}
at $\mu_R \approx m_X$.  The amount of the splitting depends on 
the explicit model as well as the distance of the conformal running, 
$m_f/m_X$.

If the splitting is very large, it leads to an extremely severe 
fine-tuning for electroweak symmetry breaking.  This will then be 
interesting (only) in the sense of Ref.~\cite{ArkaniHamed:2004fb}. 
One interesting point about obtaining the split spectrum in this 
way is that the gaugino and Higgsino masses are naturally expected 
to be the same order, $M_a \approx \mu$.

On the other hand, if the splitting is not so large, the spectrum 
does not require an extreme fine-tuning, so the scenario may be 
interesting in the context of weak scale supersymmetry.  It shows 
an interesting feature --- the gaugino masses are suppressed compared 
to the scalar masses, and yet relative values of the gaugino masses, 
as well as those of the scalar masses, exactly stay as in the standard 
gauge mediation models.  This implies that we effectively obtain 
a fractional number of messengers
\begin{equation}
  N_{\rm mess} < 1,
\label{eq:frac-Nmess}
\end{equation}
in the standard gauge mediation formula for the gaugino and scalar 
masses
\begin{equation}
  M_a = N_{\rm mess} \frac{g_a^2}{16\pi^2} 
    \frac{F_{\rm mess}}{M_{\rm mess}},
\label{eq:gmsb-Ma}
\end{equation}
\begin{equation}
  m_I^2 = 2 N_{\rm mess} \sum_a C_I^a 
    \left( \frac{g_a^2}{16\pi^2} \right)^2 
    \left| \frac{F_{\rm mess}}{M_{\rm mess}} \right|^2,
\label{eq:gmsb-mI2}
\end{equation}
where $F_{\rm mess}/M_{\rm mess} \approx \lambda(m_f) F_S/m_f$ in 
our context.  This feature of the spectrum can be tested at future 
collider experiments.

\subsection{Competition with gravity and anomaly mediation}
\label{subsec:compet}

In this subsection we discuss the competition of the gauge mediated 
contribution with both gravity and anomaly mediation.  In order for 
the predictions in the previous subsections to persist, the former 
must dominate over both of the latter.

Let us begin by estimating the size of the contributions to the 
supersymmetry breaking parameters from gravity mediation.  There 
are two classes of contributions for this: those coming from local 
operators directly connecting the hidden and SSM sector fields and 
those arising from supergravity terms (the contributions arising 
from the $F$-term VEV of the compensator field).%
\footnote{The term ``gravity mediation'' is a misnomer for the first 
 class, as it is not due to gravity.  It simply refers to contributions 
 from local operators at a scale $M_*$ of the order of the Planck 
 scale.  We, however, stick to this common terminology.}
The largest contribution for the first class typically comes from 
local operators of the form
\begin{equation}
  \int\! d^4\theta\, \frac{1}{M_*^2} S^\dagger S\, \phi^\dagger \phi,
\label{eq:magnetic}
\end{equation}
with an $O(1)$ coefficient, where $M_* \sim M_{Pl}$ is the cutoff scale.%
\footnote{If the field $S$ is an $n$-body composite operator, the cutoff 
 scale operator is suppressed by $(\Lambda_*/M_*)^{2n-2}$.  However, 
 there are lower dimension operators $\int\!d^4\theta\, Q^\dagger Q\, 
 \phi^\dagger \phi/M_*^2$, where $Q$ represents the elementary fields 
 contained in the composite field $S$.  We expect they are matched at 
 the strong scale as $Q^\dagger Q \approx S^\dagger S$.  Therefore, we 
 still obtain the operator of the size given in Eq.~(\ref{eq:magnetic}).}
If $S$ is an elementary singlet, we also have a contribution from
\begin{equation}
  \int\! d^4\theta\, \frac{1}{M_*} S\, \phi^\dagger \phi.
\label{eq:A-contr}
\end{equation}
Which of Eqs.~(\ref{eq:magnetic}) and (\ref{eq:A-contr}) gives the 
dominant contribution is then determined by the renormalization group 
scaling of these operators.

When we run Eq.~(\ref{eq:magnetic})  down to the scale $m_X$ where 
conformality is broken, we obtain
\begin{equation}
  \int\! d^4\theta\, \frac{1}{M_*^2} 
    \left( \frac{m_X}{\Lambda_*} \right)^{\alpha_S} Z_S^{-1}(m_X)\, 
    S^\dagger S\, \phi^\dagger \phi,
\label{eq:magnetic-mX}
\end{equation}
where $\alpha_S$ is the {\it same} exponent that appears in the evolution 
of the gauge mediated quadratic operators.  (If there are multiple 
exponents due to operator mixing, $\alpha_S$ is the exponent leading 
to the least amount of damping, $\hat{\alpha}_S$).  This gives 
a contribution to the supersymmetry breaking mass squared of $\phi$ 
of size
\begin{equation}
  m_{\rm grav}^2 \approx \frac{Z_S^{-1}(m_X)}{M_*^2} 
    \left( \frac{m_X}{\Lambda_*} \right)^{\alpha_S} |F_S|^2.
\label{eq:gravity}
\end{equation}
In the case that $S$ is an elementary singlet, Eq.~(\ref{eq:A-contr}) 
leads to a contribution $m_{\rm grav}^2|_{\rm sing}$, which is given 
by Eq.~(\ref{eq:gravity}) with $\alpha_S$ set to zero.

The second class of contributions for gravity mediation arises from 
supergravity terms, i.e., the $F$-term VEV of the compensator field 
$\Phi$.  This gives a contribution to the $B$ parameter of the order 
of the gravitino mass
\begin{equation}
  B_{\rm grav} \approx m_{3/2} \approx 
    \frac{|F_S|}{M_{\rm Pl}}.
\label{eq:B-grav}
\end{equation}
The other soft masses do not arise from this source (the classical 
contribution from $F_\Phi$).  However, we still need $B_{\rm grav} 
\simlt m_I$ at low energies for successful electroweak symmetry 
breaking.

Now we can compare Eqs.~(\ref{eq:gravity},~\ref{eq:B-grav}) to the 
contribution from gauge mediation.  In the case that $\alpha_S > 0$ 
(Case~1), we should compare these to the mass scale generated from 
the operators linear in $S$.  Comparing with Eq.~(\ref{eq:gravity}) 
gives
\begin{equation}
  \frac{m_{\rm grav}^2}{m_{\rm gauge}^2} 
  \approx \frac{(16\pi^2)^2}{\lambda^2}\, \frac{m_f^2}{M_*^2} 
    \left( \frac{m_X}{\Lambda_*} \right)^{\alpha_S},
\label{eq:comp-grav-1}
\end{equation}
which can easily be small if $\alpha_S$ is $O(1)$ and $\lambda$ 
is not too small (i.e., $d_S$ is not too large).  Note that the 
contribution $m_{\rm grav}^2$ is {\it suppressed}, making it more 
harmless due to the conformal dynamics.  In the case that $S$ 
is an elementary singlet ($d_S = 1$), we must also consider 
$m_{\rm grav}^2|_{\rm sing}/m_{\rm gauge}^2$.  This, however, 
can also easily be small, since $\lambda$ is then expected 
to be of order unity.

The comparison with Eq.~(\ref{eq:B-grav}), on the other hand, gives
\begin{equation}
  \frac{B_{\rm grav}^2}{m_{\rm gauge}^2} 
  \approx \frac{(16\pi^2)^2}{\lambda^2}\, \frac{m_f^2}{M_{\rm Pl}^2} 
    \left( \frac{\Lambda_*}{m_X} \right)^{2\gamma_S}.
\label{eq:comp-anom-1}
\end{equation}
This can be smaller than or of $O(1)$, i.e., the $B$ parameter is not 
too large, if $\lambda$ is not too small and $(\Lambda_*/m_X)^{2\gamma_S}$ 
not too large.  As long as Eq.~(\ref{eq:comp-anom-1}) is smaller than 
or of $O(1)$, the contribution from anomaly mediation $m_{\rm anom}^2$ 
is always subdominant to $m_{\rm gauge}^2$, since $m_{\rm anom} 
\approx B_{\rm grav}/16\pi^2$.

In the case that $\alpha_S < 0$ (Case~2), we should compare the gravity 
and anomaly mediated pieces to that coming from the operators quadratic 
in $S$.  For the contribution of Eq.~(\ref{eq:gravity}), we find
\begin{equation}
  \frac{m_{\rm grav}^2}{m_{\rm gauge}^2} 
  \approx \frac{(16\pi^2)^2}{\lambda^2}\, \frac{m_f^2}{M_*^2} 
    \left( \frac{\Lambda_*}{m_f} \right)^{|\alpha_S|}.
\label{eq:comp-grav-2}
\end{equation}
In this case the sequestering effect coming from $\alpha_S$ actually 
{\it enhances} the gravity mediated contribution relative to the 
gauge mediated contribution, and so the gravity mediated contribution 
dominates in a much larger portion of parameter space.  If $\lambda 
= O(\Lambda_*/M_*)$ (i.e., $d_S = 2$), for example, and we require 
$m_{\rm grav}^2/m_{\rm gauge}^2 \simlt 10^{-3}$, then $|\alpha_S| = 1$ 
implies that $m_f \simlt 10^{-7} \Lambda_*$.  The contribution from 
$m_{\rm grav}^2|_{\rm sing}$ is always subdominant.

For the contribution of Eq.~(\ref{eq:B-grav}), we obtain
\begin{equation}
  \frac{B_{\rm grav}^2}{m_{\rm gauge}^2} 
  \approx \frac{(16\pi^2)^2}{\lambda^2}\, \frac{m_f^2}{M_{\rm Pl}^2} 
    \left( \frac{\Lambda_*}{m_X} \right)^{2\gamma_S} 
    \left( \frac{m_X}{m_f} \right)^{|\alpha_S|}.
\label{eq:comp-anom-2}
\end{equation}
This also allows $B_{\rm grav}^2/m_{\rm gauge}^2 \simlt O(1)$. Note 
that, in contrast to $m_{\rm grav}^2$, $B_{\rm grav}^2$ does not 
have to be much smaller than the gauge mediated contribution, 
$m_{\rm gauge}^2$, since it does not contribute to flavor violation. 
Again, as long as $B_{\rm grav}^2/m_{\rm gauge}^2 \simlt O(1)$, the 
contribution from anomaly mediation $m_{\rm anom}^2$ is subdominant 
because $m_{\rm anom} \approx B_{\rm grav}/16\pi^2$.

\section{Gaugino Mediation}
\label{sec:gaugino}

An important ingredient for the solution to the $\mu$ ($B\mu$) problem 
discussed in section~\ref{subsec:mu-sol} is to have control over the 
operators of the form Eq.~(\ref{eq:A-op}), which lead to $A$ terms (as 
well as $B$ and $m_I^2$ terms).  Since these operators are not suppressed 
relative to the gaugino masses, their existence with random $O(1)$ 
coefficients would lead to large flavor violation at low energies. 
Gauge mediation allows us to have these operators under control --- 
in minimal gauge mediation (without the dynamics generating $\mu$), 
these operators are not generated at the leading order in loop or 
$F_{\rm mess}/M_{\rm mess}^2$ expansions.  We then only have to require 
that the dynamics generating $\mu$ does not induce these operators 
in such a way that they excessively violate flavor.

The argument above implies that, as long as the operators 
${\cal O}_A$ in Eq.~(\ref{eq:A-op}) are under control, the mechanism 
of section~\ref{subsec:mu-sol} can apply (not necessarily in the 
context of gauge mediation).  Interestingly, many theories in which 
the ${\cal O}_A$ operators are under control have a $B\mu$ problem 
similar to that in gauge mediation.  Consider, for example, the gaugino 
mediation scenario~\cite{Kaplan:1999ac}, in which the gauge and Higgs 
fields propagate in the bulk of an extra dimension.  The extra dimension 
is compactified on an $S^1/Z_2$ with length $L$, and the matter fields 
and hidden sector are localized on different branes.  This allows us 
to control the ${\cal O}_A$ operators.  Since the supersymmetry breaking 
field $S$ and matter fields are localized on different branes, there 
can be no direct interaction between them, including the operators of 
the form Eq.~(\ref{eq:A-op}) (taking $\phi$ to be the matter fields).

The $\mu$ and supersymmetry breaking parameters are generated 
only by the operators of the form Eqs.~(\ref{eq:scalar-op},%
~\ref{eq:Bmu-op},~\ref{eq:gaugino-op},~\ref{eq:A-op},~\ref{eq:mu-op}), 
localized on the hidden sector brane, with $\phi = H_u, H_d$.%
\footnote{We assume that the superpotential operators $W \sim 
 H_u H_d$ and $S H_u H_d$ are absent as before.}
Scaling the coefficients of these operators by naive dimensional 
analysis in higher dimensions~\cite{Chacko:1999hg}, the generated 
$\mu$ and $B\mu$ parameters are
\begin{equation}
  \mu \approx \frac{16\pi^2}{C M_* L} M_a,
\qquad
  B\mu \approx \frac{(16\pi^2)^2}{C^2 M_* L} M_a^2,
\label{eq:mu-muB-gaugino}
\end{equation}
where $M_*$ is the cutoff scale of the theory, $C$ the group theory 
factor related to the size of the gauge group, and $M_a$ the gaugino 
masses.  Now, by choosing $M_* L \approx 16\pi^2/C$, we can easily 
have $\mu \approx M_a$.  This is what we would expect if the 5D gauge 
couplings also follow naive dimensional analysis, since the 4D gauge 
couplings $g_4$ are then given by $g_4^2 \approx 16\pi^2/C M_* L 
\approx O(1)$.  However, this gives
\begin{equation}
  B = \frac{B\mu}{\mu} \approx \frac{16\pi^2}{C} M_a,
\label{eq:B-gaugino}
\end{equation}
which is too large.  The origin of this is that since the $\mu$ 
and $B\mu$ operators both contain $H_u H_d$, they are suppressed 
by the same volume factor $M_* L$.  This, however, implies that 
the suppression is canceled out in $B = B\mu/\mu$, so that $B$ is 
enhanced relative to the other soft masses.  This is analogous to 
the situation in gauge mediation where both $\mu$ and $B\mu$ are 
suppressed by the same one-loop factor.  Note that $M_* L$ must 
be larger than unity in order for the effective theory to make 
sense, so this will always enhance $B$ relative to $\mu$.

A possible solution to this problem can now be given in the same 
way as before.  Let us consider that the hidden sector becomes 
strongly interacting at the scale $\Lambda_*$, which we take to 
be close to $M_*$.  Now, if we assume that the strong conformal 
dynamics realizes Case~1 in section~\ref{sec:conseq}, then 
$B$ is suppressed relative to $\mu$ at the scale $m_X$, 
where the hidden sector leaves the conformal regime.  This 
solves the $B\mu$ problem.  The spectrum at $m_X$ is given by 
Eqs.~(\ref{eq:model-mI2}~--~\ref{eq:model-comp}).  The low-energy 
superparticle masses are then obtained by evolving these parameters 
down to the weak scale by the SSM renormalization group equations.

In order to solve the $B\mu$ problem in this way, the contribution 
of Eqs.~(\ref{eq:model-mI2}~--~\ref{eq:model-comp}) must be larger 
than or at least of the same order as the $B$ parameter arising from 
gravity mediation $B_{\rm grav} \approx m_{3/2}$.  This gives the 
condition
\begin{equation}
  \frac{B_{\rm grav}^2}{m_{\rm gaugino}^2} 
  \approx \frac{16\pi^2}{C}\, \frac{M_*^2}{M_{\rm Pl}^2} 
    \left( \frac{\Lambda_*}{m_X} \right)^{2\gamma_S} 
  \simlt O(1),
\label{eq:comp-gaugino}
\end{equation}
where we have taken $M_a \approx \sqrt{C} F_S/4\pi M_*$ at $\mu_R 
\approx \Lambda_*$, following naive dimensional analysis.  (We have 
taken the group theory factor $C$ appearing in loops to be common 
for all the fields.)  This implies that the conformal running 
distance $\Lambda_*/m_X$ cannot be large.  One application of this 
mechanism arises when the gauge groups of the standard model are 
unified into a grand unified group in the higher dimensional bulk, 
in which case successful gauge coupling unification can be preserved 
even if the compactification scale $L^{-1}$ is (slightly) below the 
conventional unification scale~\cite{Hall:2001pg}.  In this case we 
can take, for example, $C \simeq 5$ and $M_* \approx 10^{17}~{\rm GeV}$, 
which allows $\Lambda_*/m_X \approx O(10$~--~$100)$ for $\gamma_S 
\sim 0.5$, a sufficient energy interval to suppress the $B$ parameter 
(assuming that the relevant exponent $\hat{\alpha}_S$ is of order 
unity).

\section{Anomaly Mediation}
\label{sec:anomaly}

Anomaly mediation of supersymmetry breaking~\cite{Randall:1998uk,%
Giudice:1998xp} is a subtle quantum effect in which the soft supersymmetry 
breaking parameters are induced due to the superconformal anomaly.  The 
mediation is due to the presence of the $F$-component VEV of the Weyl 
compensator, which is required to cancel the cosmological constant 
once supersymmetry is broken.  The remarkable feature of anomaly 
mediation is its ultraviolet insensitivity.  Namely, no matter how 
complicated and flavor violating the theory is at high energies, once 
all supersymmetric thresholds are integrated out, the supersymmetry 
breaking effects at a given energy scale are determined only by physics 
at that energy scale, as was shown explicitly in~\cite{Giudice:1998xp,%
Boyda:2001nh}.  As a result, the flavor changing effects are virtually 
absent in the soft parameters.

For the anomaly mediated supersymmetry breaking effects to dominate, 
direct operators that couple the hidden and SSM sector fields in 
Eqs.~(\ref{eq:scalar-op},~\ref{eq:Bmu-op},~\ref{eq:gaugino-op},%
~\ref{eq:A-op},~\ref{eq:mu-op}) must be suppressed relative to the 
gravitino mass.  (In this context, we assume that the scale $M$ in 
these operators is close to $M_{\rm Pl}$.)  The original proposal 
in~\cite{Randall:1998uk} was to physically separate the two sectors 
in an extra dimension, while that in~\cite{Giudice:1998xp} was to 
require the absence of elementary singlet fields so that the operators 
Eqs.~(\ref{eq:gaugino-op},~\ref{eq:A-op},~\ref{eq:mu-op}) would 
be suppressed by simple dimensional reasons.  The motivation for 
conformal sequestering was for the purpose of suppressing the 
direct coupling operators using a four-dimensional conformal 
field theory~\cite{Luty:2001jh}.

Our new observation that the operators in Eqs.~(\ref{eq:gaugino-op},%
~\ref{eq:A-op},~\ref{eq:mu-op}) are suppressed by the wavefunction 
renormalization makes anomaly mediation possible in an even wider 
class of hidden sector models than was previously considered.  For 
example, the models of Ref.~\cite{Izawa:1996pk} have gauge singlet 
fields that acquire $F$-component VEVs, and can be made superconformal 
once a sufficient number of extra flavors is added.  If the operators 
linear in the gauge singlet fields are not sequestered, as originally 
claimed in Ref.~\cite{Dine:2004dv}, they would be dominant over the 
anomaly mediated contribution.  Especially the $A$ parameters from 
operators in Eq.~(\ref{eq:A-op}) do not respect flavor in general, 
and the resulting model would be generically excluded by the flavor 
physics data.  However, these operators actually {\it are} suppressed, 
and hence the anomaly mediated contribution dominates despite 
the presence of singlet fields in the hidden sector.

Depending on the amount of suppression, it is possible that either 
of the operators Eqs.~(\ref{eq:scalar-op},~\ref{eq:Bmu-op}) or 
Eqs.~(\ref{eq:gaugino-op},~\ref{eq:A-op},~\ref{eq:mu-op}) give 
comparable contributions to the anomaly mediated contribution. 
Suppose, for example, that the operators of Eq.~(\ref{eq:scalar-op}) 
are generated by gauge mediation and that the hidden sector shows 
the behavior of Case~2.  In this case, if the contribution from 
these operators are comparable to the anomaly mediated one, then 
the well-known problem of tachyonic sleptons in anomaly mediation 
can be solved.  In addition, in the minimal supersymmetric standard 
model (MSSM), gravity mediation gives a too large $B$ parameter: 
$B_{\rm grav} \approx m_{3/2} \approx 100~{\rm TeV}$.  This may 
be solved if the $\mu$ term of the MSSM is generated by a VEV of 
a singlet field, or if gauge mediation generates the operator of 
Eq.~(\ref{eq:Bmu-op}) at one loop, leading to a large $B$ parameter 
($\approx 16\pi^2 m_I \approx 100~{\rm TeV}$) that cancels 
$B_{\rm grav}$ at a percent level.

\section{Discussion and Conclusions}
\label{sec:concl}

In this paper, we have discussed the impact of strong hidden sector 
dynamics on the soft supersymmetry breaking parameters on general 
grounds.  While the importance of the renormalization effects on the 
operators quadratic in the hidden sector fields had been known, we have 
shown they are also important on the operators linear in the hidden 
sector fields, despite what has been stated in the literature.  This 
observation has implications both on theories of supersymmetry breaking 
and its mediation, as well as on phenomenology which may be probed 
in the near future at collider experiments.

In particular, conformal dynamics can sequester {\it both} scalar and 
gaugino masses.  However, the relative speed of sequestering is not 
calculable in general, and it is not clear which one is more important 
at the end of the conformal dynamics in a given model.  In the context 
of gauge mediation models, our result can be summarized as follows. 
If the scalar masses are suppressed faster than the gaugino masses, 
we obtain a spectrum that resembles gaugino mediation at a low 
compactification scale.  Unlike genuine gaugino mediation, however, 
there is no issue with Landau poles before reaching the unification 
scale.  In addition, $A$ terms exist, as well as Higgs soft masses 
that cancel the $\mu^2$ mass contribution.  The gravity mediated 
contribution that is potentially flavor violating is less harmful 
than in the case without the conformal dynamics.  This case also 
offers a solution to the outstanding $\mu$ ($B\mu$) problem in the 
supersymmetric standard model.  On the other hand, if the gaugino 
masses are suppressed faster than the scalar masses, the spectrum 
looks as if the ``number of messengers'' is less than unity.  In 
this case, the gravity mediated contribution is more harmful than 
in the case without the conformal dynamics.

In the context of gaugino mediation, the volume suppression factor 
tends to give $B \approx 16\pi^2 M_a$, which is unacceptable.  The 
same mechanism as in the case of gauge mediation can lead to a solution 
to the problem.  Finally, anomaly mediation may be dominant with 
conformal sequestering even if the hidden sector has a singlet field 
with gaugino mass and $A$ term operators, because they are sequestered 
as well.

We point out, however, that our analysis is limited by the lack of 
understanding of K\"ahler potential renormalizations in strongly 
coupled theories. Not only can we not work out whether the scalar 
masses or gaugino masses are sequestered more, but we could also 
worry about operators at even higher dimensions.  In this paper, 
we considered only the lowest dimension operators that can contribute 
to the soft supersymmetry breaking parameters.  However, higher 
dimension operators, such as those at cubic or quartic orders in the 
hidden sector fields, may be as important if the strong renormalization 
effects overcome the naive suppression in power counting when fields 
acquire VEVs.  Without detailed knowledge of the dynamics, we cannot 
exclude this possibility.  In addition, we assumed that the wavefunction 
renormalization factors are given solely by those in the superconformal 
limit determined by the $R$ charges.  However, realistic theories are 
necessarily perturbed by relevant operators to break supersymmetry, 
and it is possible that their impact on the wavefunction factors is 
anomalously enhanced by strong dynamics.  Note that these two issues 
are related, because one can always redefine the fields such that 
they do not acquire VEVs, but this will induce new relevant operators 
into the theory.  These effects are not possible near the Banks--Zaks 
fixed point~\cite{Banks:1981nn}, and hence are impossible to study 
using perturbation theory.

Once the LHC discovers supersymmetry, and the ILC determines the 
spectrum of superparticles precisely, it would be exciting to see 
if it shows any impact of strong hidden sector dynamics.  For this 
program, it will be important to better understand the consequence 
of strong dynamics on the renormalization of various operators, 
including higher dimension ones.  We hope that our work provides 
a step towards achieving this goal.

\section*{Note Added}

While completing this paper, we received a paper by Roy and 
Schmaltz~\cite{Roy:2007nz}.  It proposes to solve the $B\mu$ problem 
in gauge mediation using conformal dynamics of the hidden sector, 
which overlaps with our discussion in section~\ref{subsec:mu-sol}. 
They claim to obtain a spectrum identical to minimal gaugino 
mediation~\cite{Schmaltz:2000gy}, while we find that finite 
$A$ terms and Higgs soft masses that cancel the $\mu^2$ mass 
contribution are generic.

\section*{Acknowledgment}

This work was supported in part by the U.S. DOE under Contract 
DE-AC03-76SF00098, and in part by the NSF under grant PHY-04-57315. 
The work of Y.N. was also supported by the NSF under grant 
PHY-0555661, by a DOE OJI, and by an Alfred P. Sloan Research 
Foundation.  H.M. thanks Aspen Center for Physics for its support 
during his stay in summer 2007, where a part of this work was 
conducted.

\appendix

\section{Operator Mixing}
\label{app:op-mix}

In this appendix we will give some explicit examples of how different 
K\"ahler potential operators can mix with one another.  In general, 
there will be certain linear combinations of operators that evolve 
by power laws with definite exponents $\alpha$.  Some of these linear 
combinations may contain global symmetry currents of the conformal 
field theory, and will not be renormalized at all.  In the notation 
of section~\ref{sec:mech}, this means that the exponents $\alpha$ 
precisely cancel the known wavefunction renormalizations that are 
determined by the $R$ charges of the fields.

Supersymmetric $SU(N_c)$ QCD with $\frac{3}{2}N_c < N_f < 3N_c$ gives 
an unusually simple example, where $N_f$ is the number of vector-like 
flavors.  In this theory, there are four linear combinations of quadratic 
operators one can write down
\begin{equation}
  Q^\dagger Q + \bar{Q}^\dagger \bar{Q},
\qquad
  Q^\dagger Q - \bar{Q}^\dagger \bar{Q},
\qquad
  Q^\dagger T^a Q,
\qquad
  \bar{Q}^\dagger T^a \bar{Q}.
\label{eq:app-ops}
\end{equation}
The latter three correspond to the conserved $U(1)_B$, $SU(N_f)_Q$ and 
$SU(N_f)_{\bar{Q}}$ currents, and hence if these combinations appear 
in Eq.~(\ref{eq:scalar-op}), the operators are not sequestered.  On the 
other hand, the first one corresponds to the $U(1)_A$ current that is 
anomalous under the strong $SU(N_c)$ dynamics, and therefore runs with 
an exponent $\alpha_A$.  Unfortunately, we have no means to calculate 
$\alpha_A$.  In particular, we do not know whether it is positive 
or negative.

The corresponding situation in the magnetic dual 
theory~\cite{Seiberg:1994pq} is somewhat more complicated. 
In addition to the dual quarks $q$, $\bar{q}$, there are mesons 
$M$ with two indices, and hence there are many more combinations 
of operators that one can write down.  Mixing between operators 
containing the dual quarks and mesons can happen because this 
theory has the superpotential coupling ${\rm Tr}(M \bar{q} q)$.

We can classify the quadratic operators according to their 
representation under the $SU(N_f)_Q \times SU(N_f)_{\bar{Q}}$ 
symmetry of the theory.  For example, the operator 
proportional to
\begin{equation}
  {\rm Tr}(T^a M T^b M^\dagger),
\label{eq:adj-adj}
\end{equation}
transforms as (adjoint, adjoint) under the flavor group.  There are 
no other quadratic operators of the same symmetry properties, and hence 
it does not mix with any others.  It is not a conserved current and 
hence is renormalized.  Because it does not mix, it renormalizes on its 
own with a single exponent.  One cannot prove on general grounds that 
it is suppressed at low energies, but one can do explicit calculations 
close to the Banks--Zaks fixed point $N_f \approx \frac32 N_c$.%
\footnote{Up to three loops, there are no 1PI diagrams that renormalize 
 this operator, and hence the renormalization is given solely in terms 
 of the wavefunction renormalization.  There is a 1PI four-loop diagram, 
 which should generate a nonzero exponent $\alpha$, yet we do not know 
 its sign.}
Note that weakly gauging the vector-like $SU(N_f)$ flavor 
symmetry~\cite{Luty:2001jh} would still allow this operator.

The operators
\begin{equation}
  N_c {\rm Tr}(T^a M M^\dagger) + N_f (q^\dagger T^{a*} q),
\qquad
  {\rm Tr}(T^a M M^\dagger) - (q^\dagger T^{a*} q),
\label{eq:adj-sing}
\end{equation}
transform as (adjoint, singlet) under the flavor group.  The latter 
contains a conserved current, and hence is not renormalized (not 
sequestered).  The former, however, does not correspond to a symmetry 
because of the superpotential coupling, and is hence renormalized. 
Again, we do not have a general proof, but explicit calculations 
suggest that it is sequestered, with $\alpha > 0$ at the one-loop 
level.  Note that the former linear combination is the ``eigenvector'' 
of the mixing only at the one-loop level, while the precise linear 
combination is unknown at all orders.  The situation with the 
operators in which $q$'s are replaced by $\bar{q}$'s is identical.

Finally, there are three (singlet, singlet) operators
\begin{equation}
  {\rm Tr}(M M^\dagger),
\qquad
  q^\dagger q + \bar{q}^\dagger \bar{q},
\qquad
  q^\dagger q - \bar{q}^\dagger \bar{q}.
\label{eq:sing-sing}
\end{equation}
The last one corresponds to the conserved $U(1)_B$ current and hence 
is not renormalized.  The first two operators mix with unknown relative 
coefficients.  Neither of them are conserved currents and hence should 
be renormalized.  At the one-loop level, the ``eigenvectors'' of 
this mixing are
\begin{equation}
  2 {\rm Tr}(M M^\dagger) - (q^\dagger q + \bar{q}^\dagger \bar{q}),
\qquad
  N_c {\rm Tr}(M M^\dagger) + N_f (q^\dagger q + \bar{q}^\dagger \bar{q}).
\label{eq:app-eigen}
\end{equation}
The latter is sequestered already at the one-loop level with $\alpha > 0$. 
The former is accidentally conserved at the one-loop level, while it 
should receive renormalization at higher orders.  Therefore, $\alpha < 0$ 
at the lowest order for this operator.  It is not clear at all what the 
signs of the $\alpha$ exponents are in the strongly coupled situation.

The situation becomes even more complicated in theories with additional 
matter content, such as the model with an additional adjoint used in 
Ref.~\cite{Schmaltz:2006qs}.

In general, operators of the same symmetry properties mix and the 
degree of sequestering (if any) is determined by the eigenvalues of 
the mixing matrix.  Once the theory is strongly coupled, we do not 
have the techniques to work them out.  Even when the sequestering 
is plausible in theories believed to be infrared attractive, the 
signs of the exponents $\alpha$ beyond the wavefunction renormalization 
are incalculable.

\section{Generating the {\boldmath $\mu$} term in Gauge Mediation}
\label{app:mu}

In this appendix we present one simple way to generate a $\mu$ 
parameter of the same order as the gaugino masses in gauge mediation. 
We take the messenger superfields $f$ and $\bar{f}$ to transform 
under the ${\bf 10}+{\bf 10}^*$ representation of the $SU(5)_{\rm SM}$ 
symmetry containing the standard model gauge group as a subgroup, 
and introduce the superpotential interactions
\begin{equation}
  W = y f f H_u + \bar{y} \bar{f} \bar{f} H_d.
\label{eq:deltaW_mess}
\end{equation}
Here, we have imposed a $Z_2$ parity under which $f$ and $\bar{f}$ 
are odd while the other fields are even.  This has the advantage that 
mixings between the messenger and matter superfields are forbidden, 
so that the problem of flavor is not reintroduced.%
\footnote{The $Z_2$ parity makes the lightest messenger particle 
 stable, which may overclose the universe.  We can, however, simply 
 assume that the reheating temperature is low enough so that these 
 particles are not produced thermally.  Alternatively, we can (slightly) 
 modify the model.  For example, we can eliminate $Z_2$ and introduce 
 messenger matter mixings, whose sizes, however, are controlled by 
 a $U(1)$ flavor symmetry.  This modifies the third generation and 
 Higgs mass spectra (c.f. section~\ref{sec:conseq}).  Another possibility 
 is to use a messenger field that is adjoint under $SU(5)_{\rm SM}$ 
 and even under matter parity.  This allows us to avoid the introduction 
 of the flavor problem as well as the cosmological problem, without 
 an additional discrete symmetry.}
The absence of a tree level $\mu$ term is assumed.

The interactions of Eq.~(\ref{eq:deltaW_mess}) generate operators 
responsible for the $\mu$ and $B\mu$ parameters at one loop.  Integrating 
out $f, \bar{f}$ with the interactions Eq.~(\ref{eq:deltaW_mess}) 
generates
\begin{equation}
  {\cal L} = 3 \int\!d^4\theta\, 
    \frac{y \bar{y} \lambda(m_f)^*}{16\pi^2 m_f} 
     S^\dagger\, H_u H_d + {\rm h.c.},
\label{eq:app-mu-op}
\end{equation}
and
\begin{equation}
  {\cal L} = 3 \int\!d^4\theta\, 
    \frac{y \bar{y} |\lambda(m_f)|^2}{16\pi^2 m_f^2} 
     S^\dagger S\, H_u H_d + {\rm h.c.},
\label{eq:app-Bmu-op}
\end{equation}
at the scale $m_f$, where $\lambda(m_f)$ is defined in 
Eq.~(\ref{eq:lambda-mf}).  Integrating out $f, \bar{f}$ 
also generates
\begin{equation}
  {\cal L} = 3 \int\!d^4\theta\, 
    \frac{\lambda(m_f)}{16\pi^2 m_f} S 
    \left( |y|^2 H_u^\dagger H_u + |\bar{y}|^2 H_d^\dagger H_d \right) 
    + {\rm h.c.}.
\label{eq:app-AH-op}
\end{equation}
These operators contribute to $A_{H_u}$, $A_{H_d}$, $B$, $m_{H_u}^2$ 
and $m_{H_d}^2$.  Assuming $y \sim \bar{y} \sim O(1)$, these provide the 
operators discussed in section~\ref{sec:gauge}: Eqs.~(\ref{eq:gmsb-mu-op},%
~\ref{eq:gmsb-Bmu-op},~\ref{eq:gmsb-AH-op}).

\end{document}